\documentclass[11pt]{article}
\usepackage{amsmath}
\usepackage{pstricks}
\usepackage{mathrsfs}
\usepackage{youngtab}
\usepackage{bbm}

\makeatletter
\def\section{\@startsection{section}{1}{\z@}{1ex}{1ex}{\bf \large}}
\makeatother

\def\half{\frac{1}{2}}

\newfont{\bbbold}{msbm10 scaled \magstep1}

\def\bbZ{\mbox{\bbbold Z}}
\def\cA{{\cal A}}

\def\cD{{\cal D}}

\def\cF{{\cal F}}

\def\cL{{\cal L}}
\def\cM{{\cal M}}
\def\cN{{\cal N}}
\def\cO{{\cal O}}
\def\cP{{\cal P}}
\def\cQ{{\cal Q}}
\def\cR{{\cal R}}

\def\cU{{\cal U}}
\def\cV{{\cal V}}

\newfont{\goth}{eufm10 scaled \magstep1}

\def\gf{\mbox{\goth f}}
\def\gg{\mbox{\goth g}}
\def\gh{\mbox{\goth h}}

\def\d{\delta}\def\D{\Delta}

\def\F{\Phi}

\def\L{\Lambda}
\def\m{\mu}
\def\n{\nu}

\def\r{\rho}

\def\th{\theta}\def\Th{\Theta}

\def\be{\begin{equation}}\def\ee{\end{equation}}
\def\bea{\begin{eqnarray}}\def\eea{\end{eqnarray}}
\def\barr{\begin{array}}\def\earr{\end{array}}

\def\o{\omega}\def\O{\Omega}
\def\del{\partial}

\let\la=\label

\def\nn{\nonumber}
\def\bd{\begin{document}}
\def\ed{\end{document}}
\def\ba{\begin{array}}
\def\ea{\end{array}}
\def\bea{\begin{eqnarray}}
\def\eea{\end{eqnarray}}
\def\ft#1#2{\tfrac{#1}{#2}}
\def\fft#1#2{\frac{#1}{#2}}
\def\sst#1{{\scriptscriptstyle #1}}
\def\oneone{\rlap 1\mkern4mu{\rm l}}

\newcommand{\eq}[1]{(\ref{#1})}
\newcommand{\w}[1]{\\[0.#1cm]}
\def\eqs#1#2{(\ref{#1}-\ref{#2})}
\def\det{{\rm det\,}}
\def\tr{{\rm tr}}
\def\ad{{\rm ad}}

\newcommand{\hoch}[1]{$\, ^{#1}$}
\newcommand{\imperial}{\it\small Theoretical Physics Group, Imperial College London\\ Prince Consort Road, London SW7 2AZ, UK}
\newcommand{\kings}
{\it\small Department of Mathematics, King's College, University of London\\ Strand, London WC2R 2LS, UK}
\newcommand{\uu}
{\it\small Department of Theoretical Physics, Uppsala, Sweden}
\newcommand{\hip}
{\it\small HIP-Helsinki Institute of Physics, P.O. Box 64 FIN-00014
University of Helsinki, Suomi-Finland}
\newcommand{\stock}
{\it\small Department of Theoretical Physics, Stockholm, Sweden}
\newcommand{\golm}
{\it\small AEI, Max Planck Institut f\"ur Gravitationsphysik\\ Am M\"{u}hlenberg 1, D-14476 Potsdam, Germany}
\newcommand{\ihes}
{\it\small Institut des Hautes Etudes Scientifiques, 35, Route de Chartres, FR-91440 Bures-sur-Yvette, France}
\makeatletter

\newcommand{\sa}{/ \hspace{-1.2ex}}
\newcommand{\saa}{/ \hspace{-1.4ex}}
\newcommand{\saaa}{\, / \hspace{-1.6ex}}
\newcommand{\Scal}[1]{\Bigl ({#1} \Bigr )}
\newcommand{\scal}[1]{\bigl ({#1} \bigr )}

\newcommand{\CR}{\nonumber \\*}

\newcommand{\trace}{\hbox {tr}~}
\newcommand{\traceS}{\hbox {tr}_{\scriptscriptstyle \mathfrak{S}}~}

\DeclareMathAlphabet{\mathpzc}{OT1}{pzc}{m}{it}
\def\BRST{\,\mathpzc{s}\,}
\def\aBRST{{\scriptstyle (\mathpzc{s})}}
\def\q{{{\scriptscriptstyle (Q)}}}
\def\qs{{\scriptscriptstyle (Q\mathpzc{s})}}
\def\Qsla{{\mathcal{S}_{\q}}}
\def\Slav{{\mathcal{S}_\aBRST}}
\def\epsilonb{{\overline{\epsilon}}}
\def\bulletup{{\scriptstyle \bullet}}

\newcommand{\gra}[2]{{\scriptscriptstyle (#1 , #2 )}}
\newcommand{\ord}[1]{{\scriptscriptstyle (#1)}}

\def\cL{{\cal L}}
\def\cN{\mathcal{N}}
\def\cO{\mathcal{O}}

\def\ie{{\it i.e.}\ }
\def\eg{{\it e.g.}\ }

\newcommand{\sfrac}[2]{{\scriptstyle \frac{#1}{#2}}}
\newcommand{\stfrac}[2]{{\scriptscriptstyle \frac{#1}{#2}}}

 \def\balpha{{\overline{\alpha}}}
 \def\bbeta{{\overline{\beta}}}
 \def\bgamma{{\overline{\gamma}}}
 \def\bdelta{{\overline{\delta}}}
 \def\bepsilon{{\overline{\epsilon}}}
 \def\bvarepsilon{{\overline{\varepsilon}}}
 \def\bzeta{{\overline{\zeta}}}
 \def\bareta{{\overline{\eta}}}
 \def\btheta{{\overline{\theta}}}
 \def\bvartheta{{\overline{\vartheta}}}
 \def\biota{{\overline{\iota}}}
 \def\bkappa{{\overline{\kappa}}}
 \def\blambda{{\overline{\lambda}}}
 \def\bmu{{\overline{\mu}}}
 \def\bnu{{\overline{\nu}}}
 \def\bxi{{\overline{\xi}}}
 \def\bpi{{\overline{\pi}}}
 \def\brho{{\overline{\rho}}}
 \def\bvarrho{{\overline{\varrho}}}
 \def\bsigma{{\overline{\sigma}}}
 \def\bvarsigma{{\overline{\varsigma}}}
 \def\btau{{\overline{\tau}}}
 \def\bphi{{\overline{\phi}}}
 \def\bvarphi{{\overline{\varphi}}}
 \def\bchi{{\overline{\chi}}}
 \def\bpsi{{\overline{\psi}}}
 \def\bomega{{\overline{\omega}}}

\def\thalf{{\textrm{\tiny\textonehalf}}}
\def\tquarter{{\textrm{\tiny\textonequarter}}}
\def\Ko{{\scriptscriptstyle K}}
\def\tKo{\scriptscriptstyle k }
\def\corr{$\clubsuit$}

\newcommand{\auth}{\large Jesper Greitz${}^{a}$\footnote{email: jesper.greitz@nordita.org}, Paul Howe${}^{b}$\footnote{email: paul.howe@kcl.ac.uk} and Jakob Palmkvist${}^c$\footnote{email: palmkvist@ihes.fr}}

\begin{document}

\renewcommand{\thefootnote}{\fnsymbol{footnote}}

\null
\begin{flushright}

{\small NORDITA-2013-062}\\
{\small KCL-MTH-13-08}\\
{\small IHES/P/13/27}\\
{\small arXiv:1308.4972}\\

\vskip 1.5 cm
\end{flushright}

\begin{center}
{\Large{\bf The tensor hierarchy simplified}}
\vspace{.75cm}

\auth
\end{center}
\vspace{.5cm}

\centerline{${}^a${\it \small Nordita }}
\centerline{{\it \small Roslagstullsbacken 23, SE-106 91 Stockholm, Sweden }}
\vspace{.5cm}
\centerline{${}^b${\it \small Department of Mathematics, King's College London}}
\centerline{{\it \small The Strand, London WC2R 2LS, UK}}
\vspace{.5cm}
\centerline{${}^c${\it \small Institut des Hautes \'Etudes Scientifiques}}
\centerline{{\it \small 35, Route de Chartres, FR-91440 Bures-sur-Yvette, France }}

\vspace{1cm}

\centerline{{\bf Abstract}}
\vskip .5cm
\noindent A compact formulation of the field-strengths, Bianchi identities and gauge transformations for tensor hierarchies in gauged maximal supergravity theories is given. A key role in the construction is played by the recently-introduced tensor hierarchy algebra.

\vspace{1cm}


\renewcommand{\thefootnote}{\arabic{footnote}}
\setcounter{footnote}{0}

\pagebreak
\setcounter{page}{1}



It has been known for many years that the forms in $D$-dimensional  maximal supergravity theories, when the duals of the physical forms are included, are associated with algebraic structures \cite{Cremmer:1997ct,Cremmer:1998px}. These structures have been interpreted as sub-algebras of Borcherds algebras \cite{HenryLabordere:2002dk,HenryLabordere:2002xh} and in terms of extended $E$-series algebras
\cite{Julia:1997cy,West:2001as,Damour:2002cu, Riccioni:2007au,Bergshoeff:2007qi,Bergshoeff:2007vb,Riccioni:2007ni,Bergshoeff:2008xv,Riccioni:2009xr}. It has been found that there are also $(D-1)$-form potentials (de-forms), associated with deformations, and $D$-forms, otherwise known as top forms, both
carrying no physical degrees of freedom, whose existence is implied by these algebraic structures (these were first observed in $D=10$ \cite{Bergshoeff:2005ac,Bergshoeff:2006qw}). In general, the potential forms transform under representations $\cR_{\ell}$ of the duality group of the given supergravity theory where the level number $\ell$ coincides with the form-degree. In a separate, but related, development, studies of the general structure of gauged supergravities \cite{deWit:1981eq,Gunaydin:1984qu,Pernici:1984xx,Hull:1984vg,Hull:1984qz,deWit:1983gs,
Nicolai:2000sc,Nicolai:2001sv,deWit:2002vt,deWit:2003hr}
\black have revealed that the same sets of forms are needed in that context
(with two exceptions for $D=3$)  and that the gauge transformations of the potentials at level
$\ell$ involve parameters  up to level $(\ell+1)$, 
the whole set of forms giving rise to a tensor hierarchy \cite{deWit:2005hv,deWit:2008ta,deWit:2008gc}. A key feature of this general construction is the use of the embedding tensor that specifies how the gauge group $G_0$ is embedded in the duality group $G$. The embedding tensor is treated as a spurionic object that transforms under a representation of the duality group, although in a given gauging it becomes fixed and symmetry under $G$ is lost. This technique allows the formalism to be developed generally for an arbitrary gauging.

In reference \cite{Henneaux:2010ys} it was shown how one could derive Borcherds algebras for maximal supergravity theories starting from $E_{11}$, while in \cite{Palmkvist:2011vz}, it was shown how to go in  the other direction. More recently, it was argued in \cite{Kleinschmidt:2013em} that the Borcherds algebras given in \cite{HenryLabordere:2002dk,Henneaux:2010ys} for $D>7$ do not agree with those obtained by oxidation from lower dimensions. It has also become clear that the Lie superalgebras determined by the forms do not imply unique Borcherds algebras for these cases. Moreover,
these  Lie superalgebras of forms can be extended in a different way that is not symmetrical about $\ell=0$ (as the Borcherds algebras are).
The resulting new algebras, called tensor hierarchy algebras (THAs) \cite{Palmkvist:2013vya}, have the property that they encode the sequence of maps, $Y_{\ell+1,\ell}:\cR_{\ell+1}\rightarrow\cR_{\ell}$,
that appear in the formulae for the field-strengths in the  tensor  hierarchy, in a simple way,
namely as the adjoint action of a level $-1$ element corresponding to the embedding tensor. A forerunner of this type of algebra extension was given in \cite{Lavrinenko:1999xi} in the context of massive IIA supergravity where a level $-1$ element was used to describe the deformations of the field-strengths with respect to the massless case.\footnote{We are grateful to B.~Julia for pointing out this similarity.}

Borcherds algebras, extended $E$-series algebras and THAs are all
$\bbZ$-graded algebras, where
an integer $\ell\in\bbZ$ labels a non-zero subspace, and at the same time can be interpreted as the degree
of a form. 
With such an interpretation, 
these algebras are therefore truncated in a spacetime context, but in superspace there is no limit to the degree a form can have, so it is natural to include all of them \cite{Greitz:2011da}. This latter  point of view has some advantages, one of which is that the top forms can be treated gauge-covariantly because their
$(D+1)$-form field-strengths make perfectly good sense in superspace. Moreover, even in the context of on-shell maximal supergravity, there can be over-the-top forms. For example, in IIA supergravity there is a twelve-form RR field-strength tensor that has a non-zero superspace component \cite{Greitz:2011da}. This fact allows the Lie superalgebra of forms to be discussed without the complications of gauge symmetries or truncation. Moreover, it is quite possible that higher-degree forms will become non-trivial when higher-order string corrections are taken into account \cite{Greitz:2012vp}. In the context of gauging, a superspace framework allows one to discuss the complete hierarchy in a natural way without truncation \cite{Greitz:2011vh,Greitz:2012vp}.

In the current note, we shall extend the ideas of
\cite{Cremmer:1997ct,Cremmer:1998px}, taking into account the algebraic point of view of \cite{Palmkvist:2013vya}, in order to develop a simple formalism for tensor hierarchies in maximal supergravity theories, focusing on the simplest cases, $3\leq D \leq 7$. We give compact formulae for the full (infinite) sets of  field-strengths, gauge transformations and Bianchi identities. These formulae are valid in both spacetime and superspace, although, as we have mentioned, the latter framework allows one to avoid issues of truncation.
\rm For maximal supergravity theories in $3\leq D\leq7$ dimensions the forms determine a (proper) Lie superalgebra generated by the level-one elements, subject to the supersymmetry constraint, and  the duality algebra $\gg$ is simple and finite-dimensional.\footnote{In $D=8,9$,
$\gg$ is not simple and in $D=10$ IIA supergravity the Lie superalgebra of forms is not generated by the level-one forms alone. In IIB the levels are even, so the Lie superalgebra of forms is not proper (\ie has no odd elements) and in $D=11$ there is no duality group and the Lie superalgebra  has only two non-empty levels, three and six. In the last two cases there are no vectors and therefore no gaugings.
Below $D=3$ the duality groups become infinite-dimensional \cite{Julia:1982gx,Nicolai:1987kz}.}
However, the formalism can be easily adapted to higher dimensions and should be applicable in other cases such as
half-maximal supergravity theories \cite{Nicolai:2001ac,deWit:2003ja,Weidner:2006rp,Bergshoeff:2007vb,Greitz:2012vp}.
On the other hand,  it does not generalise straightforwardly to the conformal tensor hierarchies which have been studied recently in $D=6\ (1,0)$ supersymmetry 
\cite{Samtleben:2011fj,Samtleben:2012mi,Samtleben:2012fb,Bandos:2013jva,Bandos:2013sia,Palmer:2013pka}.\footnote{An underlying reason for this is that the dimensions of the forms change so that the Bianchi identities no longer define Lie superalgebras.
This is not usually seen in components since the hierarchy is truncated, but in superspace one can see that at level four one could have a cubic term in the Bianchi identity of the form $dF_5 \sim (F_2)^3 + \ldots$.}


As mentioned above, the set of forms in any maximal supergravity theory determines a Lie superalgebra, $\gf$, which is graded, not only as a superalgebra,
but also with a subspace for each positive integer $\ell$, called the level.
This can be most easily described in terms of the field-strengths $F_{\ell+1}$,
where the subscript denotes the form-degree.
At each level there will be  a set of forms determined by a representation $\cR_{\ell}$, which is generically reducible. The Bianchi identities are
\be
d F_{\ell+1}=\sum_{m+n=\ell} F_{m+1} F_{n+1}\ ,
\la{1}
\ee
while the consistency of these identities, $d^2=0$, requires, schematically
\be
\sum_{p+q+r=\ell} F_{p+1} F_{q+1} F_{r+1}=0\ ,
\la{2}
\ee
where the wedge product between the  forms is understood. These two equations determine the Lie bracket and the Jacobi identity respectively for the Lie superalgebra $\gf$. One must also require that the Bianchi identities are soluble, and one can show, using superspace cohomology, that this implies that there is a further constraint on the representations that are allowed at level two \cite{Greitz:2011vh,Greitz:2012vp}. This is the so-called supersymmetry constraint.

Let $e_{\cM}$ denote the basis elements of $\gf$ at level one, $\cM=1,\ldots, {\rm dim}\, \cR_1$. Then the basis elements at higher levels are determined sequentially by imposing the supersymmetry constraint at level two and the Jacobi identity. So at level two we have $[e_{\cM},e_{\cN}]=e_{\cM\cN}$,
and at higher levels we write
\be \label{nesting}
[e_{\cM_1},[e_{\cM_2},\ldots,[e_{\cM_{\ell-1}},e_{\cM_\ell}]\cdots]] = e_{\cM_1 \cdots \cM_\ell}\ . 
\ee
\rm 
Here and elsewhere the brackets are understood to be graded. For example, $[e_{\cM},e_{\cN}]$ is symmetric, since the level one basis elements 
$e_{\cM}$ are odd.
The level two basis elements $e_{\cM\cN}$ are then even, and the indices are projected onto the representation $\cR_2$, which is contained in the symmetric product of two $\cR_1$ representations.
However, at higher levels $e_{\cM_1\cdots \cM_{\ell}}$ will not be fully symmetric  on the indices. The notation $\langle \cM_1\cdots \cM_{\ell} \rangle$ will be used to denote the projection of $\cR_1{}^{\otimes \ell}$ onto $\cR_{\ell}$. 

We denote the Lie algebra of the duality group $G$ by $\gg$. We can obtain a new algebra $\gg_{\gf}$ by taking the semi-direct sum of $\gg$ with $\gf$. The action of $\gg$ on $e_{\cM}$ is given by
\be
[t_m,e_{\cN}]=t_{m\cN}{}^{\cP} e_{\cP}\ ,
\la{3}
\ee
where $t_m$, $m=1,\ldots, {\rm dim}\, \gg$, is a basis for $\gg$ and where $t_{m\cN}{}^{\cP}$ represents this basis in the representation $\cR_1$. Since $\gf$ is infinite-dimensional so is  $\gg_{\gf}$.

The THA 
is constructed from $\gg_{\gf}$ by appending a subspace at level $-1$, with a basis $\phi_m{}^\cM$, 
corresponding to a representation
$\cR_{-1}$ of 
$\gg$, contained in the tensor product of the adjoint and the dual representation of $\cR_1$.
The bracket of this subspace at level $-1$ with the level-one subspace is defined by
\be
[\phi_m{}^\cM,e_\cN]=\delta_\cN{}^{\lceil\cM} t_{m\rfloor}\ , \label{bracket+1-1}
\ee
where the diagonal hook brackets denote projection on $\cR_{-1}$.
The level $-1$ subspace then generates an extension of $\gg_{\gf}$  
to all negative levels,
but here we will only consider the subalgebra $\hat{\gg}$
of the THA generated by
$\gg_{\gf}$ and a single element $\Th$ at level $-1$, such that $[\Th,\Th]=0$. Thus $\hat{\gg}$ does not have any lower levels.
We set $\Th=\Th_{\cM}{}^m \phi_m{}^{\cM}$, where
we identify $\Th_{\cM}{}^m$ with the embedding tensor, a constant tensor that
describes how the
gauge group is embedded into the duality group (times a coupling strength).
The representation $\cR_{-1}$ is determined by $\cR_{2}$ since 
the Jacobi identity forces the elements
\be
[[\phi_m{}^\cM,e_{(\cN}],e_{\cP)}] - [[\phi_m{}^\cM,e_{\langle\cN}],e_{\cP\rangle}]
\ee
to vanish.
Thus the supersymmetry constraint is not only a constraint on the elements at level two, but also 
a constraint on the embedding tensor at level $-1$ (as such, it is also known as the representation constraint).
By the Jacobi identity it follows from (\ref{bracket+1-1}) that the bracket 
with $\gg$ at level zero is given by 
\be
[t_m,\phi_n{}^{\cM}]=f_{m\lfloor n}{}^p \phi_p{}^{\cM \rceil}  - t_{m \cN}{}^{\lceil \cM} \phi_{n\rfloor}{}^{\cN} \ .
\la{4}
\ee
Contracted with $\Th_\cM{}^m$, this simply says that
the embedding tensor 
transforms in the representation $\cR_{-1}$.
When one combines \eq{4} with the fact that $[\Th,\Th]=0$, one sees that the embedding tensor is invariant under the gauge 
algebra $\gg_0$ which is spanned by $X_\cM= \Th_\cM{}^m t_m = [\Th,e_\cM]$. \black

Since $\Th$ is an element at level $-1$,
its brackets with the basis elements at level $\ell$ will be at level $(\ell-1)$, \ie
\be
[e_{\cM_1\ldots \cM_{\ell}},\Th]=(-1)^{\ell}Y_{\cM_1\cdots \cM_{\ell},}{}^{\cN_{\ell-1}\cdots\cN_1 } e_{\cN_1\cdots \cN_{\ell-1} }\ ,
\la{5}
\ee
where the sign factor is included for later convenience.
Using the Jacobi identity and $[\Th,\Th]=0$  we see that $Y_{\ell+1,\ell} Y_{\ell,\ell-1}=0$, so that we can identify the $Y_{\ell+1,\ell}$ with the intertwiner that maps $\cR_{\ell+1}\rightarrow \cR_{\ell}$.
We therefore see that the THA encodes in a very concise manner
the properties of the embedding tensor and the intertwiners \cite{deWit:2005hv,deWit:2008ta,deWit:2008gc}. We refer to \cite{Palmkvist:2013vya} for a full derivation of the THA and its properties.\footnote{ The THA defined in \cite{Palmkvist:2013vya} differs from
$\hat{\gg}$ at the positive levels by the maximal ideal of $\hat{\gg}$ contained in $\gf$. This ideal corresponds to representations that are present in
the Borcherds algebra, but not seen by the tensor hierarchy, particularly a singlet and an adjoint at levels two and three for $D=3$.\black}

Let $\O$ denote the associative superalgebra of forms and $\cU_{\hat \gg}$, the enveloping algebra of $\hat\gg$.
We shall be interested in objects that take their values in the tensor product $\O\otimes\hat\gg:=\O_{\hat\gg}$, which can be viewed as a Lie superalgebra,  or in the tensor product  $\O\otimes\cU_{\hat\gg}$, which can be viewed as an associative superalgebra.
 The degree of an element in $\O_{\hat\gg}$ (or $\O\otimes\cU_{\hat\gg}$) is then the sum of the degrees of its constituents in
$\O$ and $\hat \gg$ (or $\O$ and $\cU_{\hat\gg}$).
In particular, note that odd forms anti-commute with odd elements of $\cU_{\hat\gg}$. We shall assume that the exterior derivative acts from the right (as in superspace), and also define another odd derivation acting from the right, $L_{\Th}$, that takes the bracket of a given element with $\Th$. Because $\Th$ is constant it is easy to check that
\be \label{cross-terms}
d L_{\Th}+L_{\Th} d=0\ ,
\la{6}
\ee
and as $L_{\Th}{}^2=0$ as well, it follows that the operator $d_{\Th}:=d+L_{\Th}$ is nilpotent. 

The potentials $A_\ell$, gauge parameters $\L_{\ell-1}$ and field-strengths $F_{\ell+1}$ that we consider are actually forms (with the form degrees given by the subscripts) contracted with the 
basis elements of ${\hat\gg}$ at level $\ell$. 
In order to minimise signs it is convenient to write the basis elements to the left, so for any form $\o$ at level $\ell$ we set
\be
\o=e_{\cM_{\ell}\ldots \cM_1} \o^{\cM_1\ldots \cM_{\ell}} \ .
\la{26}
\ee
We have also used here the superspace convention of summing the indices from the inside out,
although since these indices are not super
themselves, this is not really necessary.
This convention means that when we apply $d$ to a $\cU_{\hat\gg}$-valued form it starts from the right and lands directly on $\o^{\cM_1\ldots\cM_{\ell}}$.

Thus $A_\ell$ are even elements of $\O_{\hat\gg}$, while $\L_{\ell-1}$ and $F_{\ell+1}$ are
odd, and the same of course holds for their sums
\begin{align}
A&=\sum_{\ell\geq1} A_{\ell}\ , & \L&=\sum_{\ell\geq1} \L_{\ell-1}\ , & F&=\sum_{\ell\geq1} F_{\ell+1}\ . 
\end{align}
\black Note that none of these objects has a level-zero or minus one component.


We begin with the ungauged case. The formalism is essentially the same as that of \cite{Cremmer:1997ct,Cremmer:1998px}. We put
\be
F=d e^A\, e^{-A}\ . 
\la{7}
\ee
This can be considered to be a modified Maurer-Cartan form. It clearly satisfies
\be
dF+F^2=0\ .
\la{8}
\ee
Equation (\ref{8}) gives the Bianchi identities for all of the field-strength forms. These identities are consistent because the underlying algebra $\gf$ is
the Lie superalgebra of forms that was derived from the Bianchis in the first place. (Equivalently, one could view (\ref{7})
as the solution to these identities in terms of potentials.) Defining
\be
\d e^A\,e^{-A}=Z\ ,
\la{9}
\ee
we find that
\be
\d F=dZ + [Z,F] \ .
\la{10}
\ee
We want the $F$s to be gauge-invariant, so we require $dZ + [Z,F]=0$. This is solved by
\be
Z=d\L + [\L,F]\ .
\la{11}
\ee
The invariance of the $F$s then  follows straightforwardly using (\ref{8}). 

We can include the scalars in the picture in a covariant fashion by making use of the scalar fields as an element $\cV$ of the duality group $G$.
The latter acts on $\cV$ to the right globally, while the local R-symmetry group $H$ acts on the left, $\cV\rightarrow h^{-1}\cV g$. If we now set
\be
\F= d(\cV e^A)\,e^{-A}\cV^{-1}\ ,
\la{12}
\ee
then clearly $d\F+\F^2=0$.  The Maurer-Cartan form $\F$ can be rewritten as
\be
\F= d\cV \cV^{-1} + \cV F\cV^{-1}\ .
\la{13}
\ee
Now $d\cV \cV^{-1}=P+Q$, where $Q$ is the composite connection for $\gh$,  the Lie algebra of $H$, 
while $P$, which takes its values in the quotient of $\gg$ by $\gh$, can be considered as the one-form field-strength tensor for the scalar fields.
Note that $\F$ is invariant under $G$, so that we can consider $\cV F\cV^{-1}:=\tilde F$ to be the field-strength forms in the $H$-basis. The Maurer-Cartan equation for $\F$ then gives
\bea
R+ DP+ P^2&=&0\ ,\nn\w1
D\tilde F + \tilde F^2 + [\tilde F,P]&=&0\ ,
\la{14}
\eea
where $R=dQ+Q^2$ is the $\gh$-curvature and $D$ the $\gh$-covariant derivative.


The generalisation of the above to the gauged case is fairly straightforward. We define $A$ as before but then put
\be
F'=d_{\Th} e^A\, e^{-A}\ .
\la{15}
\ee
Note that $F'$ now has a level-zero component $[A_1,\Th]:=\cA$. This is the gauge field for $G_0$. So $F'=F+\cA$, where $F$ is the sum of the field-strength forms starting at level one. We have
\be
d_{\Th}F' + F'^2=0\ 
\la{16}
\ee
because $d_{\Th}{}^2=0$ as noted above in \eq{6}. As for the ungauged case we can define gauge transformations by
\be
\d e^A e^{-A}=Z\ ,
\la{17}
\ee
and if we choose
\be
Z=d_{\Th}\L -[\L_0,\Th] +[\L,F']
\la{18}
\ee
then we find, using \eq{15}, that
\bea
\d F&=&[F,[\L_0,\Th]]\ ,\la{19}\w1
\d\cA&=&[d\L_0 +[\L_0,[A_1,\Th]],\Th]\ .
\la{19.1}
\eea
The $\gg_0$-covariant  derivative on a field, \eg $\L$, for $\ell>1$, is $\cD\L=d\L + [\L,[A_1,\Th]]=d\L+[\L,\cA]$, so that \eq{19.1} is the standard formula for the gauge transformation of $\cA$ with parameter $[\L_0,\Th]$. Note that we can write
\be
Z=\cD\L + [\L_{\ell\geq2},\Th] + [F,\L]\ ,
\la{20}
\ee
or, for each level,
\be
Z_{\ell}=\cD\L_{\ell-1} + [\L_{\ell},\Th] + \sum_{m=0}^{\ell-2} [F_{\ell-m},\L_m]\ . 
\la{20.1}
\ee
The Bianchi identities, when written out, are
\be
\cD F_{\ell+1}+ (F^2)_{\ell+1} + [F_{\ell+2},\Th]=0
\la{21}
\ee
for $\ell\geq 1$. At level zero we just get the identification  of $\cF=d\cA+\cA^2$ with $-[F_2,\Th]$. These Bianchi identities are indeed what one expects for the tensor gauge hierarchy.

Expanding out \eq{15} we find for the first three $F$s
\bea
F_2&=&d A_1 + \frac{1}{2}[A_1,[A_1,\Th]] + [A_2,\Th]\ ,\nn\w1
F_3&=&\cD A_2 +  \frac{1}{2}[A_1,dA_1] + \frac{1}{3!} [A_1,[A_1,[A_1,\Th]]] \black + [A'_3,\Th]\ ,\nn\w1
F_4&=&\cD A'_3+[A_2,F_2] -\half[A_2,[A_2,\Th]] +  \frac{1}{3!}[A_1,[A_1,dA_1]]\\
&& +\frac{1}{4!}[A_1,[A_1,[A_1,[A_1,\Th]]]]\black + [A'_4,\Th] \ ,
\la{22}
\eea
where 
\bea
A'_3&=&A_3+\frac{1}{2}[A_1,A_2]\nn\ ,\w1
A'_4&=&A_4+\half[A_1,A_3]+\frac{1}{6}[A_1,[A_1,A_2]]\ .
\la{23}
\eea
These formulae, expressed in terms of the redefined gauge potentials, are the standard ones for the field-strengths in the hierarchy
\cite{deWit:2005hv,deWit:2008ta,deWit:2008gc}. The first three variations are given by
\bea
Z_1&=&\cD\L_0 + [\L_1,\Th]\nn\w1
Z_2&=&\cD\L_1 +[\L_2,\Th] + [F_2,\L_0]\nn\w1
Z_3&=&\cD\L_3 +[\L_3,\Th] + [F_3,\L_0] + [F_2,\L_1] \ .
\la{23.1}
\eea
These variations, and indeed all the  $Z_{\ell}$s in \eq{20.1},  are actually the covariant variations for the potentials, $\D A_{\ell}$, given in the literature \cite{deWit:2005hv,deWit:2008ta}. In fact they are the covariant variations for the redefined potentials $A'$ (which should be identified with the ones that are introduced from the beginning in the standard formalism).

\black
To include the scalars in the gauged case we put
\bea \label{MCf-scalars-gauged}
\F&=&\Th +d_{\Th} (\cV e^A)\,e^{-A}\cV^{-1}\nn\w1
&=&\cD\cV \cV^{-1} +\cV(\Th + F)\cV^{-1}\nn\w1
&=&\cP+\cQ +\tilde\Th+\tilde F\black \ .
\la{24}
\eea
The  extra $\Th$ term on the first line is necessary in order to obtain
$\cV\Th\cV^{-1}$ on the second line.\footnote{ This dressed version of the embedding tensor, $\tilde\Th$,  is in fact the original one, known as the T-tensor
\cite{deWit:1981eq,deWit:2002vt,deWit:2003hr}.\black}
Conjugation with $\cV$ converts $\Th$ and $F$ from the $G$-basis to the $H$-basis as indicated on the third line.
The $\cA$ gauge-field in $\cD$ acting on the scalars comes from the level-zero term in $d_{\Th}e^A e^{-A}$. It is not difficult to show that $\F$ satisfies a standard Maurer-Cartan equation $d\F+\F^2=0$. Written out it gives
\bea \label{MCeq-scalars-gauged}
R+D\cP+\cP^2&=&  - [\tilde F_2,\Th]= \cV\cF\cV^{-1}\nn\ ,\w1
D\tilde F+ \tilde F^2 +[\tilde F,\cP] + [\tilde F_{\ell\geq2},\tilde \Th]&=& 0\nn\ ,\w1
D\tilde\Th + [\tilde\Th,\cP]&=&0\ .
\la{25}
\eea
where $D=d+\cQ$ is the $\gh$-covariant derivative for the gauged theory, and $R=d\cQ +\cQ^2$.  Note that the tilded quantities do not transform under $G$ and, as a result, are also invariant under $G_0$.
 

We shall now write out a few of the above formulae in components to facilitate comparison with the literature.  For the covariant derivative one finds
(recalling \eq{26})
\be
\cD \o^{\cM_1\cdots \cM_{\ell}}=d\o^{\cM_1\ldots \cM_{\ell}}  +\big( \o^{\langle\cM_1\cdots \cM_{\ell-1} |
\cP} A_1{}^{\cN} X_{\cN\cP}{}^{|\cM_{\ell}\rangle} + (\ell-1)\ {\rm terms}\big)\ ,
\la{27}
\ee
where $X_{\cM\cN}{}^{\cP}=\Th_{\cM}{}^m t_{m\cN}{}^{\cP}$ as usual. In deriving this we have used the fact that the gauge potential $\cA$ is
\be
\cA=[A_1,\Th]=[e_{\cM} A_1{}^{\cM}, \phi_n{}^{\cN} \Th_{\cN}{}^n]=-A_1{}^{\cM}\Th_{\cM}{}^m t_m\ ,
\la{28}
\ee
where the minus sign arises in taking the odd form $A_1^{\cM}$ past the odd basis element $\phi_n{}^{\cN}$.  Using these rules one finds for the first two field-strength forms
\bea
F_2{}^{\cM}&=&d A_1{}^{\cM} +\half A_1{}^{\cN} A_1{}^\cP X_{\cP\cN}{}^{\cM} + A_2{}^{\cN\cP} Y_{\cP\cN}{}^{\cM}\ ,\nn\w1
F_3{}^{\cM\cN}&=&\cD A_2{}^{\cM\cN}-\half A_1{}^{\langle\cM}\Big(d A_1{}^{\cN\rangle}+\frac{1}{3}A_1{}^{\cP} A_1{}^{\cQ} X_{\cQ\cP}{}^{|\cN\rangle}\Big) + {A}_3'{}^{\cP\cQ\cR}Y_{\cR\cQ\cP}{}^{\cM\cN}\ .
\la{29}
\eea
For the form indices we use the superspace convention of writing the basis forms $d x^\m$ to the left, so for a $p$-form $\o$,
\be
\o=\frac{1}{p!} dx^{\m_p}\cdots dx^{\m_1}\,\o_{\m_1\cdots \m_p}\ .
\la{30}
\ee
For the field-strengths we then find, at levels one and two,
\bea
F_{\m\n}{}^{\cM}&=&2\del_{[\m} A_{\n]}{}^{\cM} -A_{[\m}{}^\cP A_{\n]}{}^{\cQ}X_{\cQ\cP}{}^{\cM} + A_{\m\n}{}^{\cP\cQ} Y_{\cQ\cP}{}^\cM\ ,\nn\w1
F_{\m\n\r}{}^{\cM\cN}&=&3\cD_{[\m} A_{\n\r]}{}^{\cM\cN}-3A_{[\m}{}^{\langle\cM} \Big(\del_{\n} A_{\r]}{}^{|\cN\rangle}-
\frac{1}{3}A_{\n}{}^\cP A_{\r]}{}^{\cQ}X_{\cQ\cP}{}^{|\cN\rangle} \Big)+\nn\w1
&\phantom{=}& + \ {A'}_{\m\n\r}{}^{\cP\cQ\cR} Y_{\cR\cQ\cP}{}^{\cM\cN}\ .
\la{31}
\eea
The minus signs in the middle terms are due to the superspace summation convention. To get superspace formulae one simply has to substitute super-indices, $M,N$, etc, running over both $x$ and $\th$ coordinates, for $\m,\n$, etc. 

In summary, the field-strengths for the tensor hierarchy in gauged maximal supergravity theories are given by the generalised Maurer-Cartan form \eq{15}, the Bianchi identities by the generalised Maurer-Cartan equation \eq{16} and the covariant gauge transformations by $Z$ defined in \eq{17}. In more detail, the gauge transformations and the Bianchi identities are given by \eq{20} and \eq{21} respectively, while the expressions for the field-strengths have to be extracted from the general definition \eq{15} level by level. The first three are given in \eq{22}. These are the standard expressions as we confirmed by writing the first two out after removing the basis elements in \eq{29}. Finally, we note that the scalars can also be incorporated in a manifestly covariant fashion by including them as a superspace (or spacetime)-dependent element $\cV$ of the duality group in the Maurer-Cartan form $\F$.

\subsection*{Acknowledgments}
JP would like to thank B.~Julia, S.~Lavau and H.~Samtleben for discussions.

\raggedright

\end{document}